# A Review on the Applications of Crowdsourcing in Human Pathology

Roshanak Alialy[1], MD, Sasan Tavakkol[2*], MSc, Elham Tavakkol[3], MD, Amir Ghorbani-Aghbologhi[4], MD, Alireza Ghaffarieh[5], MD, Seon Ho Kim[2], PhD, Cyrus Shahabi[2], PhD

1- Department of Medicine, David Geffen School of Medicine, University of California, Los Angeles, CA, USA.
2- Department of Computer Science, Viterbi School of Engineering, University of Southern California, Los Angeles, CA, USA.
3- Telemedicine Research Center, National Research Institute of Tuberculosis and Lung Diseases (NRITLD), Shahid Beheshti University of Medical Sciences, Tehran, Iran
4- Department of Pathology and Laboratory Medicine, University of California, Davis Medical Center, Sacramento, CA, USA.
5- Department of Pathology and Laboratory Medicine, Indiana University School of Medicine, Indianapolis, IN, USA.

## ABSTRACT

The advent of the digital pathology has introduced new avenues of diagnostic medicine. Among them, crowdsourcing has attracted researchers' attention in the recent years, allowing them to engage thousands of untrained individuals in research and diagnosis. While there exist several articles in this regard, prior works have not collectively documented them. We, therefore, aim to review the applications of crowdsourcing in human pathology in a semi-systematic manner. We firstly, introduce a novel method to do a systematic search of the literature. Utilizing this method, we, then, collect hundreds of articles and screen them against a pre-defined set of criteria. Furthermore, we crowdsource part of the screening process, to examine another potential application of crowdsourcing. Finally, we review the selected articles and characterize the prior uses of crowdsourcing in pathology.

## INTRODUCTION

In 2006 the term "crowdsourcing" was coined by Howe [1] to describe an open call for outsourcing tasks, usually done by an employee or an agent, to a large group of people. Although this term was introduced only a decade ago, the concept was utilized centuries ago in 1714, when the Britain's Parliament offered a prize in the form of an open call to find a method to identify the longitudinal position of a ship [2]. The advent of the Internet and prevalence of mobile devices provided a convenient channel for reaching out the crowd and recruiting participants and boosted the growth of crowdsourcing applications. Today, the term crowdsourcing is usually used for only online open calls, and several platforms exist which offer services for this purpose. In this paper, we review the articles which have used crowdsourcing in a human pathology related study.

Crowdsourcing is now in widespread use, covering a varied domain of studies from classifying galaxies [3], to increase situational awareness for disaster response and recovery [4]. Applications of crowdsourcing are not limited to controlled researches, for example, the Ushahidi-Haiti project (i.e., Mission "4636") gathered more than 80,000 text messages from on-site users regarding the local situation and needs after the 2010 Haiti earthquake. A large number of these texts were translated into English using a crowdsourcing approach and were sent to the first responders to perform emergency activities [5].

[*] Corresponding Author (tavakkol@usc.edu)

Health research and medical science have increasingly utilized crowdsourcing in the recent years. In 2013, Ranard et al. [6] reviewed the applications of crowdsourcing in the health research. They identified four types of crowdsourcing tasks in the reviewed articles: problem-solving, data processing, surveillance/monitoring, and surveying. They concluded that crowdsourcing could effectively improve the quality, decrease the cost, and increase the speed of a project. They also discussed that current efforts lack a standardized guideline to collected well-defined metrics, which hurts the clarity and comparability of methodologies.

Good and Su [7], explored the bioinformatics community and studied how crowdsourcing is utilized in different contexts. They divided crowdsourcing tasks into two major groups: microtasks, and megatasks. They classified methods found in the literature for crowdsourcing microtasks by the incentives used to recruit the participants: altruism for volunteers (i.e., citizen scientists), fun in casual games (i.e., games with a purpose [8]), money in commercial platforms (e.g., Amazon Mechanical Turk), obligation in workflow sequestration (e.g., ReCAPTCHA [9]), and learning in education (i.e., incorporating crowdsourcing in the curriculum of courses [10]). Regarding megatasks, authors mentioned two possible forms: hard games (e.g., Foldit [11] and EteRNA [12]), and open innovation contests (e.g., DREAM challenges [13]). They distinguish casual games and hard games according to the incentive for the gamers, which is engagement in repetitive tasks (referred as 'grinding') for casual games, and engagement with a difficult challenge in hard games.

The rich literature regarding applications of crowdsourcing in health and medical sciences indicates that it is a promising tool for researchers. However, each field of medicine has its unique requirements, limitations, and opportunities. In this study, we aim to review the applications of crowdsourcing only in pathology. We define a method to conduct a systematic search of the literature and then filter the articles step by step similar to the recommended procedure in a systematic review.

# BACKGROUND

A 60% drop in academic pathology in the UK between 2000 and 2012 [14], and a predicted 35% drop in the ratio of pathologists to the population in US between 2010 and 2030 [15] are evidences that pathology is facing a severe lack of workforce resources [14]. Automated methods and machine learning are promising methods to automate routine scoring and evaluations [16]–[20]. However, reducing the workload for pathologists, and development and training of these methods require a large amount of validated data sets which are currently not widely available [14]. Crowdsourcing, not only may generate the required data for training these algorithms, but also can be effectively used to outsource tedious but relatively simple tasks to the crowd, especially in low-resourced areas. For instance, a pathologist is required to check 1000 of red blood cell (RBC) images for an accurate malaria diagnosis. This becomes a severe bottle-neck considering the fact that majority of Malaria cases are reported in low-resourced locations, such as in Africa [21]. Crowdsourcing can be used to filter out most of the RBC images (i.e., those which are obviously clean/infected) and leave the experts with a handful of images to make a final diagnosis. Crowdsourcing can also be helpful in pathology research, where thousands of tissue samples must be scored. In this case, since the results do not directly affect patients, researchers can tolerate a higher but known level of error introduced by the crowd and consider it in their results.

In 2013, Ranard et al. [6] performed a systematic review on three major literature search engines, namely PubMed, Embase, and CINAHL, exploring the applications of crowdsourcing in health research. At that time, they found only 21 articles meeting their criteria, which was surprisingly small. Among the selected articles, only three were classified as pathology related researches. Inspired by this prior work, we aim to review applications of crowdsourcing specifically in human pathology. Since the prior review in 2013, encompassed pathology research, our review is beneficial only if there is a substantial number of relevant articles published after 2013. To assess this issue, we searched the mentioned three search engines using the same Boolean expression as Ranard et al. and inspected the number of returned results over time. Figure 1 shows that the cumulative number of articles returned from search engines for the Boolean expression used by Ranard et al. has experienced an exponential growth. Results show that 1126, 1105, and 112 were returned from PubMed, Embase, and CINAHL, respectively in 2016, compared to only 265, 343, and 36 in 2013. This almost three-fold increase clearly necessitates conducting a new review on this subject. However, to limit our scope of review, we only consider pathology related articles from the broader health and medicine research topic.

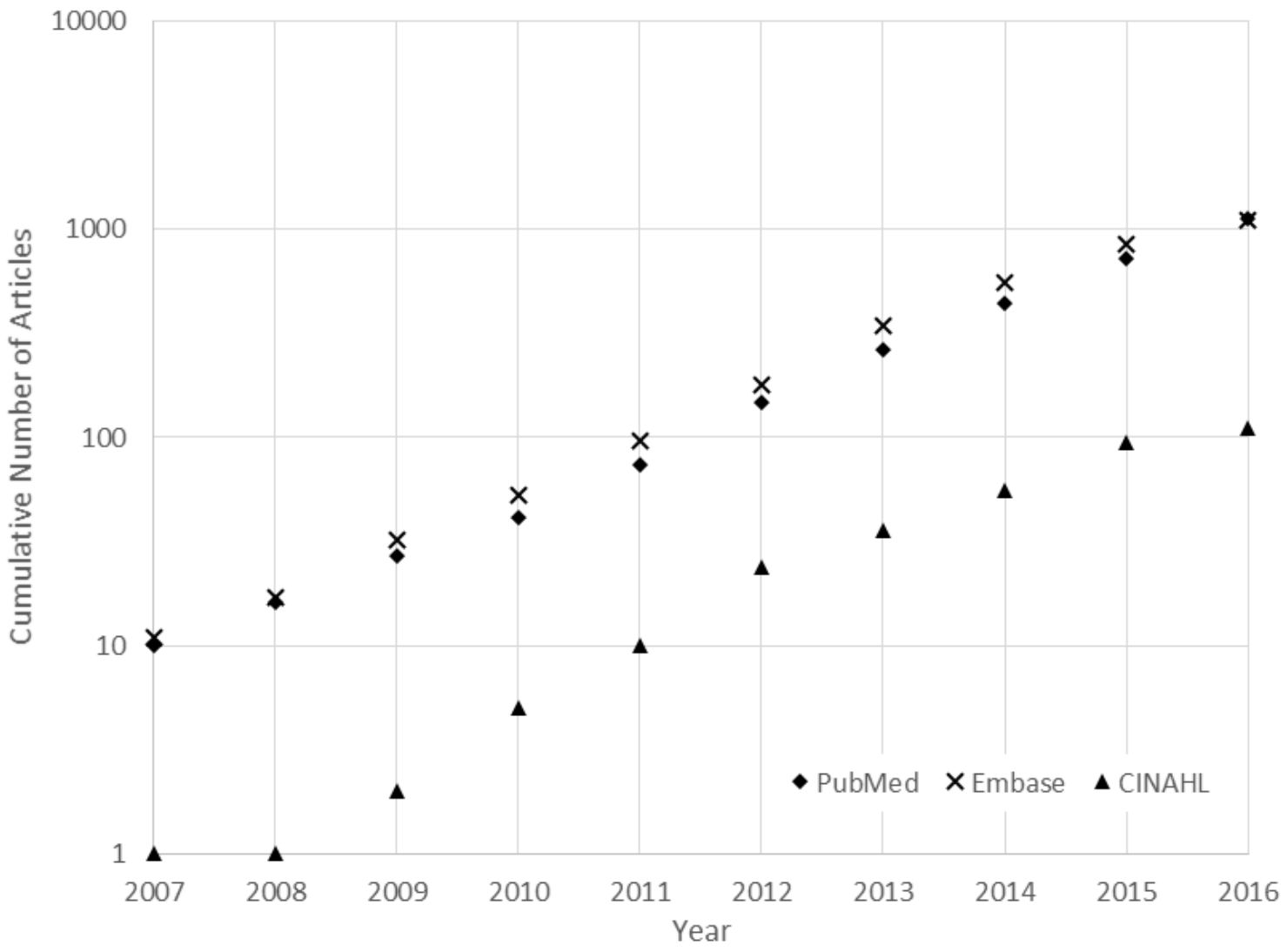

*Figure 1 – Cumulative number of articles returned from different search engines for Boolean expression used in [6] for crowdsourcing and citizen science research.*

# METHODOLOGY

## Definitions

*Crowdsourcing*: We define crowdsourcing as an online open call to outsource a defined task to the crowd. Our definition, therefore, excludes surveying the crowd for information about themselves (e.g., their own general health) and experts soliciting opinions from other experts. We also exclude competitions and challenges, where participants are most likely teams of experts. It must be noted that our definition suits the purpose of this review and is not recommended to be used otherwise.

*Pathology*: We define pathology according to its MeSH term definition: “A specialty concerned with the nature and cause of disease as expressed by changes in cellular or tissue structure and function caused by the disease process [22].” Furthermore, we only consider articles related to human pathology.

## Search Strategy

Following Ranard et al. [6], we decided to search PubMed, Embase, and CINAHL for the relevant literature. To perform a systematic search, we needed to carefully construct a Boolean expression which would return articles likely to be related to both crowdsourcing and pathology. To find literature related to crowdsourcing we use the same Boolean string defined by Ranard et al.:

cr_expression = 'crowdsourc* OR "crowd source" OR "crowd sourcing" OR "crowd sourced" OR "citizen science" OR "citizen scientist" OR "citizen scientists"' Eq. 1

However, to find the results related to pathology, we used a more rigorous approach to define a Boolean expression. It is important to note that adding "OR (pathology)" to *cr_expression* from Eq. 1, is not sufficient, as there might be crowdsourcing papers which are closely related to the field of pathology but the term "pathology" does not appear in their abstracts or titles. Therefore, we expand our Boolean expression by adding terms which their inclusion in an article is likely to be an indication of relevance to pathology.

Let's define P(*pathology*|*w*), as the probability that a document containing word *w* is related to pathology. We want to rank all the candidate words per this value and include those with relatively higher ranks in our search Boolean. To calculate this probability for each a word, using the Bayes' theorem is helpful:

$$\mathrm{P}(pathology|w) = \frac{\mathrm{P}(w|pathology)}{\mathrm{P}(w|pathology) + \mathrm{P}(w|not_pathology)}\mathrm{P}(pathology) \quad \text{Eq. 2}$$

where P(*w*|*pathology*) is the probability that *w* exists in an article related to pathology, P(*w*|*not_pathology*) is the probability that *w* exists in an article not related to pathology, and P(*pathology*) is the probability that a given article is related to pathology. Since P(*pathology*) is constant for all the candidate words, its value does not change our ranking, and therefore we drop this value and define P'(*pathology*|*w*) as:

$$\mathrm{P}'(pathology|w) = \frac{\mathrm{P}(w|pathology)}{\mathrm{P}(w|pathology) + \mathrm{P}(w|not_pathology)} \quad \text{Eq. 3}$$

We need text corpuses of article abstracts to extract a set of candidate words and calculate the probabilities in Eq. 3 for this set. Let's call the text corpus of pathology related articles as the P-Corpus, and the corpus containing articles not related to pathology as the N-Corpus. With this definition, we have:

$$\mathrm{P}(w|pathology) = \frac{n_{w,P}}{N_P} \quad \text{Eq. 4}$$

$$\mathrm{P}(w|not_pathology) = \frac{n_{w,N}}{N_N} \quad \text{Eq. 5}$$

where $n_{w,P}$ and $n_{w,N}$ are the number of documents containing *w* in the P-Corpus and N-Corpus, respectively. $N_p$ and $N_N$ are the total numbers of documents in the P-Corpus and N-Corpus, respectively. We can now select the top *n* words with larger values of P'(*pathology*|$w_i$) and construct a Boolean expression for searching pathology related articles as:

p_expression = '$w_1$ OR $w_2$ OR $w_3$ OR … $w_n$' Eq. 6

To limit our results to those articles which meet the conditions imposed by both *cr_expression* and *p_expression* we combine them with an AND:

final_expression = (cr_expression) AND (p_expression) Eq. 7

We used *final_expression* as defined by Eq. 7 for our searches in PubMed, Embase, and CINAHL.

### Selection Criteria

Articles returned for the constructed Boolean expression were included in our review only if they met all the conditions defined below:

1. Majorly written in English.
2. Representing a primary research (i.e., not a review paper, editorial note, etc.).
3. Published in a peer-reviewed journal.
4. Not a duplicate article.
5. Related to human pathology.
6. Using crowdsourcing as part of the study.

Excluded papers therefore were, those not meeting at least one of the criteria mentioned above, those which were falsely filtered out (false negatives), those which did not contain any of the keywords used in our Boolean expression in the title or abstract.

### Literature Informed Search

As the final step in our search strategy, we performed a literature informed search, based on the articles which met our selection criteria. In this step, we reviewed the references of each selected article, articles authored by the first author of a selected article, and articles cited a selected article. We then included those which met our selection criteria in this study.

### Extracted Information

We reviewed articles which passed our multi-stage screening and extracted following information from each one: research field, study objective, study outcome, crowdsourced task, length of crowdsourcing, whether a tutorial was used, crowdsourcing platform, task load per participant, recruiting and advertisement channel, whether a monetary incentive was used, data validation and performance assessment technique, whether a screening step was included, total task size, crowd size, crowd age, crowd geographic location. We also looked for other crowd demographics such as ethnicity, language, education, etc. However, none of the selected articles included this information.

## RESULTS

### Search Terms Selection

We developed Python tools to implement the search strategy explained in the previous sections. We utilized the Entrez Programming Utilities to query PubMed and build our two corpuses. To build our P-Corpus, we downloaded all the 172,723 abstracts published in 2016 and returned from PubMed for the search term “pathology”. To build the N-Corpus, we needed to query PubMed and download the abstracts which do not contain the word “pathology”. However, since at least one inclusive word must be present in the PubMed search Boolean, we added the generic term “university” to our expression and used “university NOT (pathology)” as the Boolean search term. Furthermore, we downloaded only the first 1000 results returned for each month in 2016, in this case. This strategy avoided downloading an unnecessarily large number of articles and ensured that we have a uniform sample of abstracts over the year. We made a text corpus of these 12,000 abstracts and used it as our N-Corpus.

Before calculating the probability values, we removed all the stop-words for the English language present in the Natural Language Toolkit (NLTK) Python package from both corpuses. Stop-words are the most common words in a language (e.g., “the”, “at”, “is”, etc. for English). These words are often removed in some natural language processing techniques (e.g., bag-of-words) as their presence in a text document usually do not give us any useful information. The probability of finding a stop-word in a document is almost certain and therefore the information conveyed but this presence is close to zero. Since our final goal was deriving search terms, we did not use stemming. Stemming is a process where words are reduced to their word stem or root. For example, words such as “pathology”, “pathologies”, and “pathological” can all be reduced to a common root such as “patholog”. We did not find this technique to be useful for our purpose.

We calculated the probability values for 403,929 words found in the P-Corpus and 96,675 words found in the N-Corpus. Table 1 shows the top 15 words and the values of P(*w*|*pathology*) or P(*w*|*not_pathology*) for them. As can be seen, the most frequent words in each corpus, are not informative. This is expected according to the definition of P(*w*|*pathology*) or P(*w*|*not_pathology*). However, we will show that ranking words per their P(*pathology*|*w*) values results in a meaningful set of words.

*Table 1: Ten most probable words in P-Corpus and N-Corpus*

| w (P-Corpus) | P(w\|pathology) | w (N-Corpus) | P(w\|not_pathology) |
|---|---|---|---|
| pmid | 0.985 | pmid | 0.980 |
| author | 0.964 | university | 0.979 |
| information | 0.964 | information | 0.977 |
| doi | 0.945 | author | 0.977 |
| department | 0.797 | Doi | 0.863 |
| university | 0.778 | department | 0.651 |
| epub | 0.683 | pmcid | 0.571 |
| medline | 0.615 | pmc | 0.571 |
| indexed | 0.614 | epub | 0.544 |
| pathology | 0.576 | results | 0.444 |

We filtered out all the words with P(*w*|*pathology*) < 0.05 and considered the rest of the words in the P-Corpus as the candidate words (434 words in total). We then, calculated the value of P(*pathology*|*w*) for these candidates and included those with P(*pathology*|*w*) > 0.70 in our set of “pathology related words”. This resulted in 61 words. However, we reviewed these words and manually removed 18 words, which we believed are not closely enough related to pathology, despite ranking high by our methodology (e.g., “diagnostic”, “therapeutic”, “expression”, etc.). We also removed 5 plural words “cells”, “tumors”, “tissues", “genes”, and “mice” and only kept their singular forms. Table 2 shows the final 38 pathology related words that we included in our Boolean search. We constructed our *p_expression* using the words given in this table. Figure 2 illustrates this set in a words cloud.

*Table 2: Selected set of pathology related words.*

| *w* | $P(\mathit{pathology} \mid w)$ | *w* | $P(\mathit{pathology} \mid w)$ | *w* | $P(\mathit{pathology} \mid w)$ |
|---|---|---|---|---|---|
| pathology | 0.999 | signaling | 0.818 | vivo | 0.751 |
| pathologist* | 0.966 | mouse | 0.817 | inflammatory | 0.751 |
| neoplasm* | 0.904 | progression | 0.810 | cellular | 0.741 |
| biopsy | 0.890 | lesions | 0.786 | tissue | 0.738 |
| histology* | 0.889 | pathway | 0.781 | pathways | 0.738 |
| metastasis | 0.885 | inflammation | 0.769 | lung | 0.730 |
| malignant | 0.860 | inhibition | 0.765 | liver | 0.729 |
| carcinoma | 0.857 | oncol | 0.765 | gene | 0.728 |
| tumor | 0.839 | breast | 0.761 | molecular | 0.728 |
| apoptosis | 0.834 | cell | 0.760 | protein | 0.726 |
| immunology | 0.833 | receptor | 0.760 | genetics | 0.715 |
| proliferation | 0.833 | cancer | 0.758 | mol | 0.711 |
| prognostic | 0.827 | oncology | 0.755 | vitro | 0.706 |
| prognosis | 0.824 | immune | 0.752 | | |

* $P(w|\mathit{pathology})$ for this word was less than 0.05, but keyword is added to the list as suggested by the anonymous reviewer.

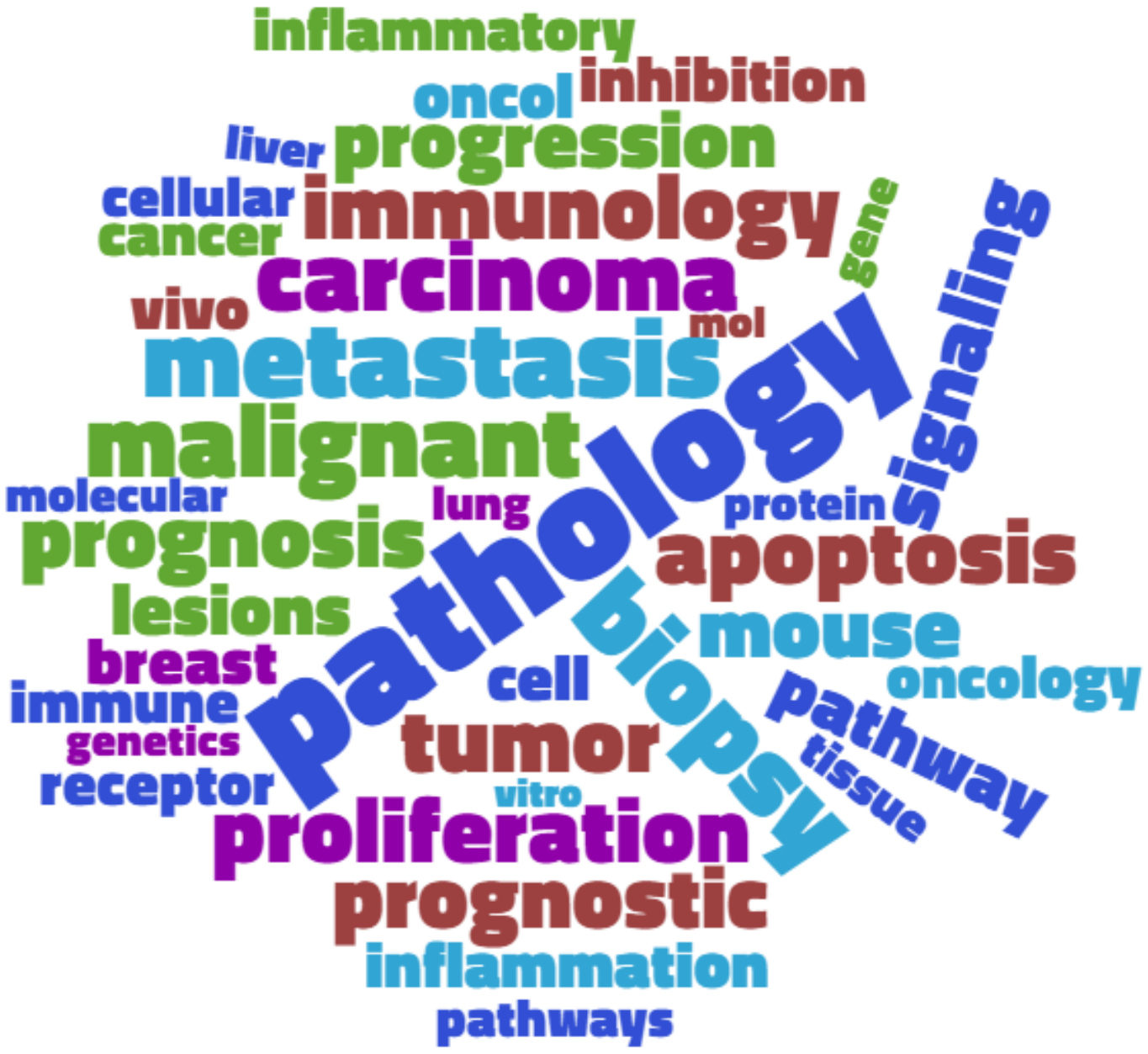

*Figure 2 – Word cloud of the pathology related words. Size of each world shows how likely it will return a pathology related article.*

## Search Results

We searched PubMed, Embase, and CINAHL on April 14, 2017, using our Boolean search string, *final_expression*, as defined by Eq. 7. This resulted in collecting 726 articles, which were reduced to 487 after removing the duplicates. We then performed a multi-stage screening to identify those articles which met our selection criteria. In the first step, we removed non-primary articles (i.e., abstracts, editorials, reviews, etc.), articles in other languages, and those which were obviously not pathology or health related. The first screening step left 129 articles in our cohort.

In the second step, we crowdsourced the review process through an online platform (Qualtrics.com). A simple question was asked like: "Is this article related to human pathology: *a crowd-sourcing approach for the construction of …*? (1) Yes, (2) Not Sure, (3) No". The title of the article was hyperlinked to the abstract of the paper on the publisher's website. Questions were presented to the participants in a random order. We placed a screening question in the beginning of the experiment: "Are you a medical doctor/student? (1) Yes, (2) No", and only the participants who answered "Yes" were able to continue. A second question asked if the participant is a pathologist or a pathology resident. Those who reviewed more than 15 papers, were promised to be acknowledged in our paper. All participants agreed to use their answers in this study.

We recruited participants through social media namely Facebook, LinkedIn and Twitter. The experiment was live during May 2017. A total of 31 participants passed our screening question and 4 participants claimed to be pathologists/pathology residents. The maximum and minimum numbers of reviewed article per participant were 105 and 1, respectively. A total number of 640 answers were collected, where 244, 116, and 280 answers were "Yes", "Not Sure", and "No", respectively. We ignored "Not Sure" answers, as if the participant did not see the corresponding question. The maximum number of collected answers for a single article was 8 and the minimum number was 2. There were 5 articles with only 2 answers: a "Yes" and a "No". We directly included them in the next screening stage for further review, as it was not possible to derive a conclusion based on the crowdsourcing results in those cases. Total of 68 Articles with 50% or more "Yes" answers qualified to the next stage. We finally, added 10 articles to the final screening stage, for which there were less than 50% of "Yes" answers, but at least 1 pathologist has marked the articles as related to pathology. Therefore, a total number of 83 articles passed our crowdsourcing screening stage.

In the final screening step, we read the full texts of 83 remaining articles to select those which met our selection criteria. The focus in this step was identifying articles which use crowdsourcing as defined in our study and are related to human pathology. Only six articles met all our selection criteria at the end. The literature informed searched added one more article to this set. Figure 3 shows the selection flow in a Sankey diagram[†]. We reviewed the final seven articles [14], [21], [23]–[27] and extracted their information which is discussed in the next section.

† Sankey Diagram Generator by Dénes Csala, based on the Sankey plugin for D3 by Mike Bostock; https://sankey.csaladen.es; 2014

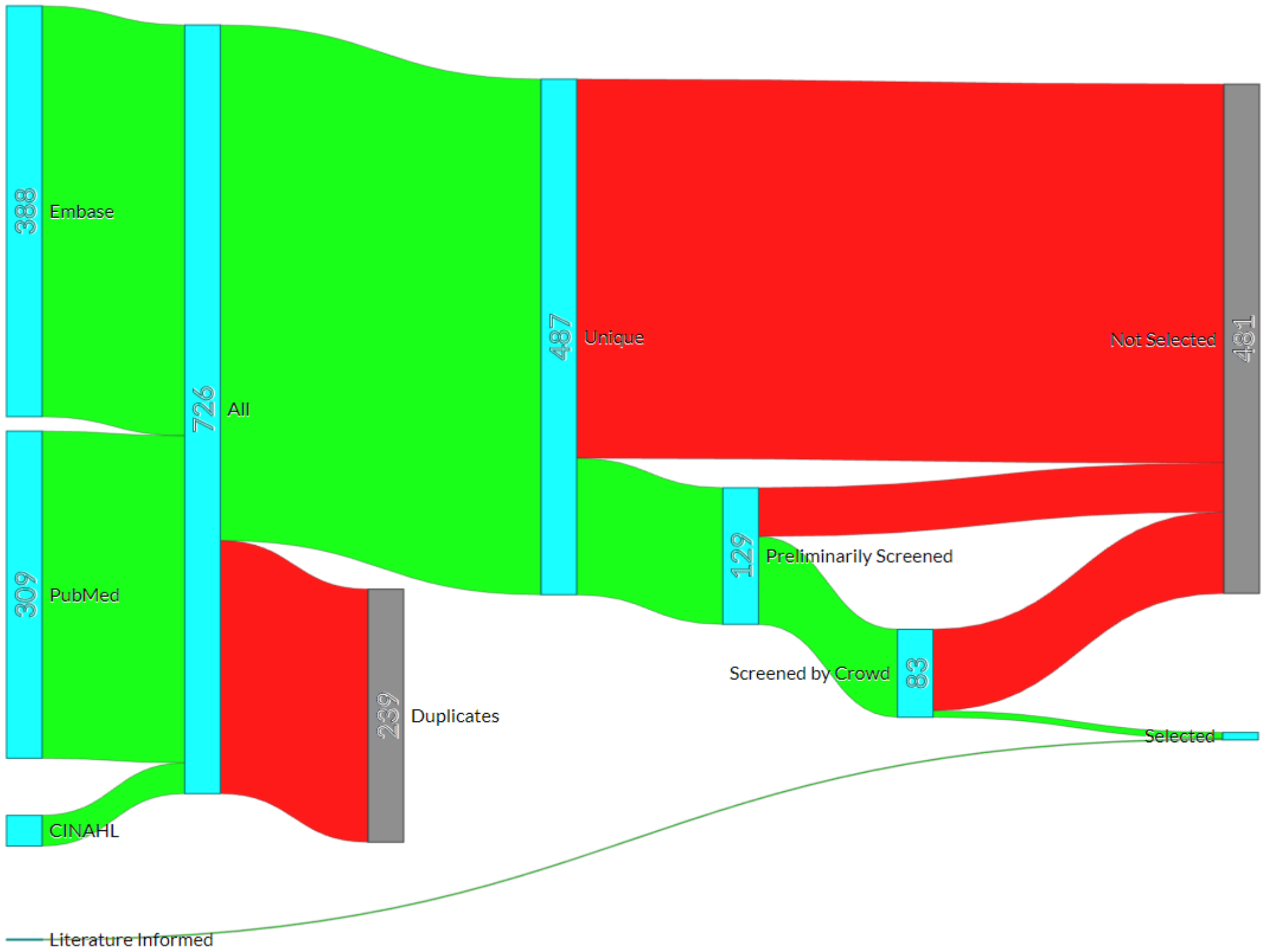

*Figure 3 – Sankey diagram of the selection flow. Red represents "filtered out" flow and green represents "filtered in" flow.*

## Reviewed Articles

The crowdsourced task in all the articles in our cohort was image analysis. Among the selected seven articles, three were related to malaria diagnoses [21], [23], [24], and four to cancer diagnoses [14], [25]–[27]. Four topics were covered by the articles: microbiology, immunohistochemistry, molecular pathology, and in-vivo microscopy. Table 3 summarizes the scope of each article.

### Microbiology

Three of seven articles were classified under the topic microbiology [21], [23], [24]. All three articles were related to malaria diagnosis in red blood cell (RBC) images and all used a custom game as the crowdsourcing platform. Mavandadi et al. [21] designed a game to let untrained gamers identify malaria infected RBCs in images under light microscopes. They demonstrated with 31 of participants that crowdsourcing via games can achieve the accuracy of medical experts in making diagnoses. Later, they scaled up their experiment [23] with over 2000 of untrained gamers and achieved the same conclusion as before. In an independent experiment, Luengo-Oroz et al. [24] et al. tested whether untrained volunteers can count malaria parasites in digitized images of thick blood smears. They conducted experiment via a Web-based game and over 6000 participants, concluding that nonexpert players may achieve an accuracy higher than 99 % in a parasite counting.

### Molecular Pathology

Molecular pathology covered only one article [25]. Authors in this study let the public score tumor images and concluded that crowdsourcing is a viable method for tumor scoring. They also shared their complete data set under the Creative Commons license [28] which can be invaluable for other researchers to conduct further research on the results.

## In-vivo Microscopy

One of the articles [26] was a crowdsourcing research on the relatively new field of in-vivo microscopy [29]. The researchers in this article studied whether untrained crowd could rapidly be trained to accurately distinguish between cancer and benign tissues. They achieved a high level of diagnostic accuracy, validating they hypothesis.

## Immunohistochemistry

Immunohistochemistry was the main topic of two papers. In the first article [14], researchers showed that crowdsourcing is sufficiently accurate to detect cancer cells and IHC biomarker scoring. In the second article [27], authors found crowdsourcing as a promising approach for scoring protein expression in IHC images and for large scale cancer molecular pathology studies in Immunohistochemistry. They further compared the results achieved by crowdsourcing to automated methods and showed that using crowdsourcing obtains a better concordance with the pathologist interpretations.

*Table 3: Articles summary.*

| Title | Crowdsourcing task | Research Field | Study Objective | Study Output |
|---|---|---|---|---|
| “Distributed medical image analysis and diagnosis through crowd-sourced games: a malaria case study.” [21] | Identify malaria infected red blood cells (RBCs) imaged under light microscopes. | Microbiology | Whether untrained humans can be reliable in microscopic analysis of biomedical samples toward diagnosis [of Malaria]. | Crowdsourcing via games can achieve the accuracy of medical experts in making diagnoses as "an accuracy within 1.25% of the diagnostics decisions made by a trained medical professional" was achieved. |
| “Crowd-sourced biogames: managing the big data problem for next-generation lab-on-a-chip platforms.”[23] | Mark malaria infected and healthy cells in RBC images. | Microbiology | Scale up a prior small experiment to determine whether crowdsourcing via a large-scale online game could identify RBCs infected with malaria parasites. | Previously untrained gamers can make diagnoses, achieving an accuracy of 98.13% compared to the ground truth data. |
| “Crowdsourcing Malaria Parasite Quantification: An Online Game for Analyzing Images of Infected Thick Blood Smears.” [24] | Tag as many parasites as possible in an RBC image in 1 minute. | Microbiology | Whether anonymous previously untrained volunteers can count malaria parasites in digitized images of thick blood smears via a Web-based game. | By combining player inputs, “nonexpert players achieved a parasite counting accuracy higher than 99%”. |
| “Crowdsourcing the general public for large scale molecular pathology studies in cancer.” [25] | Identify if cancer cells are present in an image and estimate the number of cancer cells. Also, estimate the proportion of nuclei staining positive and its intensity. | Molecular Pathology | Share images of tumors with the general public to score tumor markers. | Crowdsourcing is a viable method for tumor scoring and provides accurate ER data. The area under ROC curve was as high as 0.95 for cancer cell identification and 0.97 for ER status. |
| "Optical biopsy of bladder cancer using crowd-sourced assessment." [26] | Assess CLE video sequences derived from benign (n = 3) or cancerous (n = 9) urothelium. | In-vivo Microscopy | Whether "non-medically trained crowd could learn to rapidly and accurately distinguish between cancer and benign tissues." | "The crowd achieved an overall diagnostic accuracy of 92% for cancer classification and exceeded 70% accuracy for cellular borders, vasculature, and cellular morphology." |
| "Crowdsourcing for translational research: analysis of biomarker expression using cancer microarrays." [14] | Identify regions where cancer was present in a sequence of images. | Immuno-histochemistry | Study whether crowdsourcing is sufficiently accurate to detect cancer cells and IHC biomarker scoring, and if tutorials enhance crowd's performance. | High Spearman correlations between lay participants and experts for immunohistochemistry was achieved. Short training improved accuracy of detecting cancer in samples. |
| "Crowdsourcing scoring of immunohistochemistry images: evaluating performance of the crowd and an automated computational method." [27] | Estimates the percentage of cancer positive/negative nuclei stained in colors and label the image. | Immuno-histochemistry | Quantify IHC images for labeling of ER status with image-labeling and nuclei-labeling, then compare their performance with automated methods. | Crowdsourcing is a promising approach for scoring protein expression in IHC images and for large scale cancer molecular pathology studies in Immunohistochemistry, obtaining a better concordance with the pathologist interpretations than automated method. |

## Logistics of Crowdsourcing

Four articles reported the length of active crowdsourcing period which ranged from 9.5 hours [26] to 20 months [25]. Six articles reported having a tutorial to train the participants on the subject before performing the required task. One article, specifically studied the effect of tutorial on the performance of the crowd and reported an increase in the accuracy of detecting cancer in images [14]. Three articles used custom games to gamify the crowdsourcing process, and the rest designed a website using open source or commercial frameworks.

Five articles reported the method of recruiting, where CrowdFlower and Amazon Mechanical Turk (AMT) each were used in one article. Two studies used newsletter, blogs, and social media (e.g., Facebook). One article recruited its participants, using direct emails. In three cases, monetary incentives were given to participants, and one article claimed that participants receiving monetary incentives are more reliable [14].

All seven articles validated the crowdsourcing results by comparison to experts' opinion. Two studies used previously scored images to give feedback to the participants during performing the task. This enabled participants to evaluate themselves on the go and possibly improve their performance. Five articles used a method for screening the participants and included the results only from those who passed the screening. For example, one article required participants to score at least 99% on the training session to be able to continue [21]. Table 4 summarizes the logistics of crowdsourcing in all seven studies.

| *Table 4: Logistics of Crowdsourcing* | | | | | | | | |
|---|---|---|---|---|---|---|---|---|
| Ref. | Length of crowd sourcing (days) | Tutorial | Platform | Task Size | Channel | Monetary incentives (USD) | Data validation and performance assessing techniques | screening |
| [21] | N/A | Each player was shown a brief online tutorial which explained the rules of the game and "how malaria infected RBCs typically look with some example images." | Custom game | we had a total of 7829 characterized human RBC images, | N/A | No | Roughly 20% of all the images in the game were control images scored by experts, and were used to give feed-back to the player. | Players were required to successfully complete a training game with at least 99% accuracy to continue. |
| [23] | 3 months | A brief online tutorial on what malaria infection looks like. | Custom game | 8500 individual RBC images , collectively diagnosed more than 1 million RBCs | N/A | No | Known controls images ranked by medical experts were used to calculate individuals performance. The answers were weighted accordingly. | Only players who rated at least 100 images were included in results. |
| [24] | 1 month | The splash screen of the game briefly explained what is a parasite and what it is not. | Custom game | The image database was compiled from 28 Giemsa-stained thick films made from blood infected with malaria (Plasmodium falciparum) parasites | Facebook (30%) and Twitter(30%) directed the major traffic. The rest come "through links in digital newspapers and blogs, especially from Spanish-speaking countries." | No | "Each click was compared with the gold standard" and player could immediately see if s/he made a correct or incorrect selection. | A player should have found all the parastises in an image to move on to the next image. |
| [25] | 20 months | The task and key steps required to score each image were described in a brief web based training tutorial | Custom website | 180,172 sub-images derived from images of 12,326 tissue microarray cores labeled for estrogen receptor (ER). A total | Media outlets such as Huffington Post and the UK terrestrial television channel ITV which covered the project. A post on Reddit.com and | No | A set of 200 sub-images were scored by a pathologist and was used as a standard to | No |
| | | upon entry of the participant. | | of 1,939,984 sub-image classifications were available for analysis | advertising on Facebook resulted in a huge spike in the participants number. | | calculate the user performance score in identifying cancer cells in images. | |
| [26] | 9 hours and 27 minutes | Crowd workers completed a training module. | Custom software. | CLE video sequence randomly selected from a set of 12 sequences derived from a benign (n = 3) or cancerous (n = 9) urothelium (Figure | Amazon Mechanical Turk | 50¢ for each video assessed. | Videos were previously annotated by an expert user. | Yes |
| [14] | N/A | A 10 to 15 min basic tutorial consisted of passive, text and image-based set of instructions, and was developed based on interviews and training sessions with pathologists. | Trailblazer releases were developed using an online crowdsourcing open-source framework, Pybossa (https://github.com/PyBossa/). | Ten images were overlaid by a 6 6 grid for a total of 360 squares (Figure 1) for Detecting cancer cells. We selected 21 lung/EGFR cytoplasmic stain samples and 30 bladder/p53 nuclear stain samples representative of most clinical samples to test this, whereby each participant scored a random set of 10 images | Recruiting was done "through e-mails to individuals registered for non-pathology Cancer Research UK crowdsourcing projects." New volunteers were also recruited for Trailblazer through newsletters and advertising. The Prolific Academic platform was also used to recruit additional paid testers. | £7.5 per hour for paid participants | Experts scored images were as ground truth and for validation | No |
| [27] | N/A | N/A | CrowdFlower platform was used to design crowdsourcing applications. | 380 images in the pilot study + 5,338 image labeling + 5,338 image for nuclie lableing | CrowdFlower | The platform charged $282 for the image labeling (4 hours) job and $2,280 for nuclei labeling job (472 hours). No info on the participants compensation. | Results were compared with expert pathologists' opinion for validation. | Yes |

### Crowd Characteristics

All seven papers reported the crowd size, however in one case, only the number of participants in sub-experiments were reported. The crowd sized varied from only 31 to almost 100,000. The average task load was reported in three articles and ranged from 6 images to 9394 images per user. Age range of the crowd was reported in one article and geographic location in two. No further information on the demographics of the crowd were available. The characteristics of the crowd in each article is shown in Table 5.

*Table 5: Characteristics of the Crowd*

| Ref. | Crowd Size | Task load per user | Age | Geographic Location |
|---|---|---|---|---|
| [21] | 31 unique participants (non-experts), | Each player scored all the 9394 images which took, on average, less than one hour for each player. | ranging between the ages of 18 and 40 | N/A |
| [23] | 2,150 | N/A | N/A | 77 countries |
| [24] | more than 6000 | N/A | N/A | 95 countries shown on a map. |
| [25] | 98,293 | Median of 6 images per user | N/A | N/A |
| [26] | 602 crowd workers were | N/A | N/A | N/A |
| [14] | Only for some sub-experiments the number of participants was given. | 10 images | N/A | N/A |
| [27] | Image Labeling 113 155 61 52<br>Nuclei Labeling 3,244 1,572 2,216 1,243 | N/A | N/A | N/A |

## DISCUSSION

To the best of our knowledge, this is the first study to review applications of crowdsourcing in human pathology. The number of articles which applied crowdsourcing in a pathology related study and met our other selection criteria (journal article, etc.) was surprisingly small at only seven. Despite widespread application of crowdsourcing in practical situations in other research areas, all the reviewed articles in this study applied crowdsourcing in a research setting. While medical applications of crowdsourcing need special attention, considering the possible privacy issues and the level of required accuracy, it is about time to see non-research applications of crowdsourcing in pathology. It is important to note that our results do not include all the possible articles. Gray literature, conference papers, articles not indexed in our selected search engines are among those articles which are ignored in our study.

Researchers (Mavandadi et al.) suggested that one way to involve untrained participants in making medical diagnoses, is to let them filter the data for medical expert. For example, in resource-poor areas, RBC images can be reviewed by a large crowd and an expert can only view those samples with higher uncertainty in results. Therefore, crowdsourcing in pathology needs a generic framework to efficiently combine crowds' and experts' opinions. Tavakkol et al. [30] introduced a framework based on the entropy of crowdsourced messages to maximize the information gain under limited interpretation resources. Such a framework can be adapted in medical and pathology researches.

Demographics of the crowd such as age, education, language, etc. may have significant effect on the results and therefore must be an essential part of any study. However, among the seven reviewed articles only two articles reported information, though minimally, about the demographics of the crowd. This finding is in agreement with Ranard et al [6]. This might be because of the limitations in the current available crowdsourcing platforms and channels, where most of the time demographics of the crowd are not available and thus cannot be stored.

All the reviewed articles concluded that their initial hypothesis was valid and crowdsourcing is a viable approach to perform certain assumed tasks. While these studies help promoting crowdsourcing in pathology, they do not help understanding the limitations of this approach. For instance, an essential question in using crowdsourcing in diagnoses is the possible complications that may occur regarding patient's privacy. Furthermore, tasks which need highly skills experts cannot be crowdsourced, but further studies are necessary to determine the threshold where crowdsourcing is not an option anymore.

As expected from the nature of the research field, the main crowdsourced task in all the selected articles was image analysis. However, several types of images such as gross, histological, cytological, immunohistochemical, etc. are used for diagnosis in pathology. This variety, raises a question regarding the effects of type of imagery in the performance and limitations of crowdsourcing. For example, the analyzing gross images may not be as entertaining as stained images for the crowd, which in turn may limit our ability to gamify this process. Furthermore, difficulty level of detecting abnormalities in images may change according to the type of the image, making crowdsourcing more suitable for a certain type. Since each of the selected articles studied only a certain type of imagery, further studies are necessary to perform a comparative analysis on the influence of type of pathological images on the logistics and performance of crowdsourcing.

One of the reviewed articles [27] compared the results of crowdsourcing with automated methods and concluded that the crowd did a better job. However, computer vision techniques and digital pathology are continuously experiencing advances. The question is, if there will be a time that automated methods will surpass the crowd in every aspect, and what the estimated time for such an event might be. Furthermore, if automated methods eventually, and in the next few years, will outperform the crowd in all pathology related tasks, is there any reason to keep crowdsourcing in our toolbox? Some researchers suggest that crowdsourcing can be used to generate data for training automated methods. Furthermore, researchers believe that there is a social impact in crowdsourcing which cannot be ignored. For example, [25] engaged about 100,000 individuals in scoring breast cancer tissue samples, which inevitably increases the awareness of this crowd regarding a global issue of breast cancer.

## CONCLUSION

We introduced a novel method to effectively construct a Boolean expression for systematic literature search. This method can be used by other researchers in systematic reviews, decreasing the subjectivity of defining a Boolean expression. Utilizing this methodology, we collected potentially related articles to applications of crowdsourcing in human pathology. Furthermore, we, ourselves, used crowdsourcing to review the collected articles and decrease the number of potential articles. After screening the candidates against our selection criteria, we only found seven articles. We believe that the pathology is well-suited for outsourcing some of the reparative but simple tasks to the crowd, and therefore such a small number of relevant papers was surprising. Finally, we observed that all the related articles used the crowd for image analysis in one of the following areas: microbiology, molecular pathology, in-vivo microscopy, and immunohistochemistry.

## ACKNOWLEDGEMENTS

We specially thank medical doctors Fardad Masoumi and Scott Jafarian who were generous with their times and reviewed over 100 articles in our crowdsourcing stage. We also thank medical doctors Amir Mohammad Pirmoazen, Mohammad Nargesi, Amir Alishahi Tabriz, Negar Haghighi, Ali Gharibi, Aditya Keerthi Rayapureddy, Faridokht Khorram, Nelli S. Lakis, and Marco Vergine for their assistance in reviewing articles and participating in our crowdsourcing open call.

## REFERENCES

[1] J. Howe, "The rise of crowdsourcing," *Wired Mag.*, vol. 14, no. 6, pp. 1–4, 2006.

[2] D. Sobel, Longitude: The true story of a lone genius who solved the greatest scientific problem of his time. Bloomsbury Publishing USA, 2007.

[3] C. Lintott, K. Schawinski, S. Bamford, A. Slosar, K. Land, D. Thomas, E. Edmondson, K. Masters, R. C. Nichol, M. J. Raddick, and others, "Galaxy Zoo 1: data release of morphological classifications for nearly 900 000 galaxies," *Mon. Not. R. Astron. Soc.*, vol. 410, no. 1, pp. 166–178, 2010.

[4] H. To, S. Tavakkol, S. H. Kim, and C. Shahabi, "On Acquisition and Analysis of Visual Data for Crowdsourcing Disaster Response," 2016.

[5] N. Morrow, N. Mock, A. Papendieck, and N. Kocmich, "Independent evaluation of the Ushahidi Haiti project," *Dev. Inf. Syst. Int.*, vol. 8, p. 2011, 2011.

[6] B. L. Ranard, Y. P. Ha, Z. F. Meisel, D. A. Asch, S. S. Hill, L. B. Becker, A. K. Seymour, and R. M. Merchant, "Crowdsourcing--harnessing the masses to advance health and medicine, a systematic review.," *J. Gen. Intern. Med.*, vol. 29, no. 1, pp. 187–203, 2014.

[7] B. M. Good and A. I. Su, "Crowdsourcing for bioinformatics," *Bioinformatics*, vol. 29, no. 16, pp. 1925–1933, 2013.

[8] L. Von Ahn and L. Dabbish, "Designing games with a purpose," *Commun. ACM*, vol. 51, no. 8, pp. 58–67, 2008.

[9] L. Von Ahn, B. Maurer, C. McMillen, D. Abraham, and M. Blum, "recaptcha: Human-based character recognition via web security measures," *Science (80-. ).*, vol. 321, no. 5895, pp. 1465–1468, 2008.

[10] P. Hingamp, C. Brochier, E. Talla, D. Gautheret, D. Thieffry, and C. Herrmann, "Metagenome annotation using a distributed grid of undergraduate students," *PLoS Biol.*, vol. 6, no. 11, p. e296, 2008.

[11] S. Cooper, F. Khatib, A. Treuille, J. Barbero, J. Lee, M. Beenen, A. Leaver-Fay, D. Baker, Z. Popović, and others, "Predicting protein structures with a multiplayer online game," *Nature*, vol. 466, no. 7307, pp. 756–760, 2010.

[12] J. Lee, W. Kladwang, M. Lee, D. Cantu, M. Azizyan, H. Kim, A. Limpaecher, S. Gaikwad, S. Yoon, A. Treuille, and others, "RNA design rules from a massive open laboratory," *Proc. Natl. Acad. Sci.*, vol. 111, no. 6, pp. 2122–2127, 2014.

[13] D. Marbach, J. C. Costello, R. Küffner, N. M. Vega, R. J. Prill, D. M. Camacho, K. R. Allison, M. Kellis, J. J. Collins, G. Stolovitzky, and others, "Wisdom of crowds for robust gene network inference," *Nat. Methods*, vol. 9, no. 8, pp. 796–804, 2012.

[14] J. Lawson, R. J. Robinson-Vyas, J. P. McQuillan, A. Paterson, S. Christie, M. Kidza-Griffiths, L.-A. McDuffus, K. A. Moutasim, E. C. Shaw, A. E. Kiltie, and others, "Crowdsourcing for translational research: analysis of biomarker expression using cancer microarrays," *Br. J. Cancer*, vol. 116, no. 2, p. 237, 2017.

[15] S. J. Robboy, S. Weintraub, A. E. Horvath, B. W. Jensen, C. B. Alexander, E. P. Fody, J. M. Crawford, J. R. Clark, J. Cantor-Weinberg, M. G. Joshi, and others, “Pathologist workforce in the United States: I. Development of a predictive model to examine factors influencing supply,” *Arch. Pathol. Lab. Med.*, vol. 137, no. 12, pp. 1723–1732, 2013.

[16] A. BenTaieb, M. S. Nosrati, H. Li-Chang, D. Huntsman, and G. Hamarneh, “Clinically-inspired automatic classification of ovarian carcinoma subtypes,” *J. Pathol. Inform.*, vol. 7, 2016.

[17] K. Oikawa, A. Saito, T. Kiyuna, H. P. Graf, E. Cosatto, and M. Kuroda, “Pathological diagnosis of gastric cancers with a novel computerized analysis system,” *J. Pathol. Inform.*, vol. 8, 2017.

[18] Y. Bar, I. Diamant, L. Wolf, and H. Greenspan, “Deep learning with non-medical training used for chest pathology identification,” in *Proc. SPIE*, 2015, vol. 9414, p. 94140V.

[19] A. Janowczyk and A. Madabhushi, “Deep learning for digital pathology image analysis: A comprehensive tutorial with selected use cases,” *J. Pathol. Inform.*, vol. 7, 2016.

[20] E. A. Mohammed, M. M. A. Mohamed, B. H. Far, and C. Naugler, “Peripheral blood smear image analysis: A comprehensive review,” *J. Pathol. Inform.*, vol. 5, 2014.

[21] S. Mavandadi, S. Dimitrov, S. Feng, F. Yu, U. Sikora, O. Yaglidere, S. Padmanabhan, K. Nielsen, and A. Ozcan, “Distributed medical image analysis and diagnosis through crowd-sourced games: a malaria case study,” *PLoS One*, vol. 7, no. 5, p. e37245, 2012.

[22] “Pathology - MeSH,” *National Center for Biotechnology Information*. [Online]. Available: https://www.ncbi.nlm.nih.gov/mesh/68010336.

[23] S. Mavandadi, S. Dimitrov, S. Feng, F. Yu, R. Yu, U. Sikora, and A. Ozcan, “Crowd-sourced BioGames: managing the big data problem for next-generation lab-on-a-chip platforms,” *Lab Chip*, vol. 12, no. 20, pp. 4102–4106, 2012.

[24] M. A. Luengo-Oroz, A. Arranz, and J. Frean, “Crowdsourcing malaria parasite quantification: an online game for analyzing images of infected thick blood smears,” *J. Med. Internet Res.*, vol. 14, no. 6, 2012.

[25] F. J. C. dos Reis, S. Lynn, H. R. Ali, D. Eccles, A. Hanby, E. Provenzano, C. Caldas, W. J. Howat, L.-A. McDuffus, B. Liu, and others, “Crowdsourcing the general public for large scale molecular pathology studies in cancer,” *EBioMedicine*, vol. 2, no. 7, pp. 681–689, 2015.

[26] S. P. Chen, S. Kirsch, D. V Zlatev, T. Chang, B. Comstock, T. S. Lendvay, and J. C. Liao, “Optical biopsy of bladder cancer using crowd-sourced assessment,” *JAMA Surg.*, vol. 151, no. 1, pp. 90–93, 2016.

[27] H. Irshad, E.-Y. Oh, D. Schmolze, L. M. Quintana, L. Collins, R. M. Tamimi, and A. H. Beck, “Crowdsourcing scoring of immunohistochemistry images: Evaluating Performance of the Crowd and an Automated Computational Method,” *Sci. Rep.*, vol. 7, 2017.

[28] F. J. C. dos Reis, S. Lynn, H. R. Ali, D. Eccles, A. Hanby, E. Provenzano, C. Caldas, W. J. Howat, L.-A. McDuffus, B. Liu, and Others, “Research data supporting ‘Crowdsourcing the General Public for Large Scale Molecular Pathology Studies in Cancer.’” [Online]. Available: http://www.repository.cam.ac.uk/handle/1810/247569.

[29] Y. Chen, C.-P. Liang, Y. Liu, A. H. Fischer, A V Parwani, and L. Pantanowitz, “Review of advanced imaging techniques,” *J. Pathol. Inform.*, vol. 3, 2012.

[30] S. Tavakkol, H. To, S. H. Kim, P. Lynett, and C. Shahabi, “An entropy-based framework for efficient post-disaster assessment based on crowdsourced data,” in *Proceedings of the Second ACM SIGSPATIALInternational Workshop on the Use of GIS in Emergency Management*, 2016, p. 13.